\documentclass[%
 reprint,
superscriptaddress,
 amsmath,amssymb,
 aps,
cuted,
]{revtex4-2}

\usepackage{graphicx}
\usepackage{dcolumn}
\usepackage{bm}
\usepackage[dvipsnames]{xcolor}
\usepackage{tikz}
\usepackage{pgfplots}
\usepackage{subcaption}
\usepackage{ulem}
\newcommand{\tphi}{\tilde{\phi}}
\newcommand{\Vd}{V_0}

\pgfplotsset{compat=1.17}
\begin{document}

\title{Substrate Curvature Curbs the Coffee Ring Effect}

\author{John McCarthy}
\affiliation{%
Department of Engineering Science, University of Oxford, Parks Road, OX1 3PJ, Oxford, United Kingdom}
\author{Alfonso Castrej\'on-Pita}
\email{alfonso.castrejon-pita@wadham.ox.ac.uk}
\affiliation{%
Department of Engineering Science, University of Oxford, Parks Road, OX1 3PJ, Oxford, United Kingdom}
\author{Mokhtar Adda-Bedia}
\affiliation{%
Universit\'e de Lyon, Ecole Normale Sup\'erieure de Lyon, CNRS, Laboratoire de Physique, F-69342 Lyon, France.}
\author{Dominic Vella}
\email{dominic.vella@maths.ox.ac.uk}
\affiliation{%
Mathematical Institute, University of Oxford, Woodstock Rd, Oxford, OX2 6GG, UK.}

\date{\today}

\begin{abstract}
When the liquid phase of a particle-laden droplet evaporates, a ring of solute is typically formed --- what has become known as the `coffee ring effect'. A key focus of recent work has been the suppression of the coffee-ring effect to leave behind more spatially-uniform particle coatings instead. Efforts to suppress the coffee ring effect often focus on physical effects such as Marangoni flows. Here we focus on a purely geometric effect --- the effect of substrate curvature --- by evaporating suspension droplets on spherical surfaces of different radius of curvature. We find that stains formed on more highly curved  substrates are more uniform. To understand this, we propose a model of the evaporation-induced flow, combined with a detailed calculation of how curvature modifies the local evaporation rate. This model shows that evaporation in the centre of the droplet is enhanced, leading to increased concentration of solute there and a reduction in the propensity for ring-formation. 
\end{abstract}

\maketitle

The evaporation of liquid films containing a suspension is a common way of forming solid coatings in settings from  painting in everyday life to spin coating. Evaporation of suspension droplets is of particular interest  not only because of the ubiquity of the phenomenon, but also the wealth of related applications from inkjet printing to diagnostics \cite{lohse2020,lohse2022,mampallil2015}. The purpose of evaporating droplets is usually to produce a uniform coating, but faces a key impediment in the `coffee ring effect' (CRE)\cite{deegan1997}: drying suspension droplets tend to produce a ring-like stain in which the solute is concentrated at the edge, while the centre is depleted. This inhomogeneity is caused by the pinning of the contact line  during evaporation, which, in turn, drives fluid from the centre of the droplet to the edge. This outwards radial flow also advects solute to the contact line where it is deposited to form a ring. The radial flow and hence the CRE, is also enhanced by the presence of a singularity in the evaporative flux at the contact line, but takes place even when evaporation occurs uniformly over the droplet's surface \cite{deegan1997,moore2021}.

Since the mechanism leading to the CRE relies on contact line pinning, a natural strategy to control it is to reduce pinning. This has been tested using liquid infused surfaces (LIS) and hydrophobic textured surfaces \cite{kuang2014, dicuangco2014}. The contact-line of a droplet on a LIS is free to move, but a consequence of this is that the deposits formed by evaporation on LIS tend to have tight clusters in the centre of the contact region, rather than a uniform layer of deposit that is the goal of most coating approaches  \cite{kuang2014}. Methods that do not rely on increasing contact line mobility have also been shown to be effective. The coffee-ring can be prevented by capturing particles at the interface using standing acoustic waves \cite{mampallil2015}, accelerating evaporation \cite{li2016} or altering the shape of the particles \cite{yunker2011}. The Marangoni effect has also been utilized to reverse the direction of the radial flow inside the droplet and thereby suppress the CRE \cite{hu2006}. Other strategies to reduce the CRE target the evaporative flux itself, rather than the flow that it induces.
 For example, droplets placed in sealed containers \cite{kajiya2008} or of particularly high concentration \cite{kim2018} have different evaporative fluxes and, as a result, reduced ring-formation. Similarly a change in droplet and/or substrate geometry can play a role: non-axisymmetric contact lines affect the resulting stain  \cite{saenz2017,moore2021} while placing a droplet in a well \cite{dambrisio2021}, on a pillar \cite{li2019} or on a cylinder \cite{paul2021} alters the evaporation dynamics and hence the deposit formed. 

That both droplet morphology and substrate geometry modify the evaporation rate (including the strength of the evaporative singularity) can be understood from a molecular perspective as follows. The rate of evaporation of solute  is determined by molecular diffusion; modifying the number/density of neighbouring molecules therefore modifies the concentration gradient and hence the evaporation rate. As discussed above, this fundamental mechanism has been exploited to control the formation of solutal rings. In this Letter we show that substrate curvature also gives control over the CRE, whilst preserving the axisymmetry of the resulting stain.

To study the effect that substrate curvature has on deposit form, we performed a number of experiments. Droplets of a nigrosine-water solution (0.1\% w/w) (with initial volumes $\Vd$ ranging from  $2-5\mathrm{~\mu L}$) were deposited on Delrin spheres with radii ranging from 1-10 mm. Once the droplet had evaporated and a stain had been produced, the sphere was photographed. (To ensure uniform lighting, the sphere was placed inside a tube of diffusion plastic for photography --- this was surrounded by a flexible LED panel, as shown in Fig.~1(a).) The contact area and average radius of the stain were extracted from these images and used to calculate the angular position of the contact line, $\phi_c$, defined in Fig.\ 1(a).   

\begin{figure}
\centering
\includegraphics[width= \linewidth]{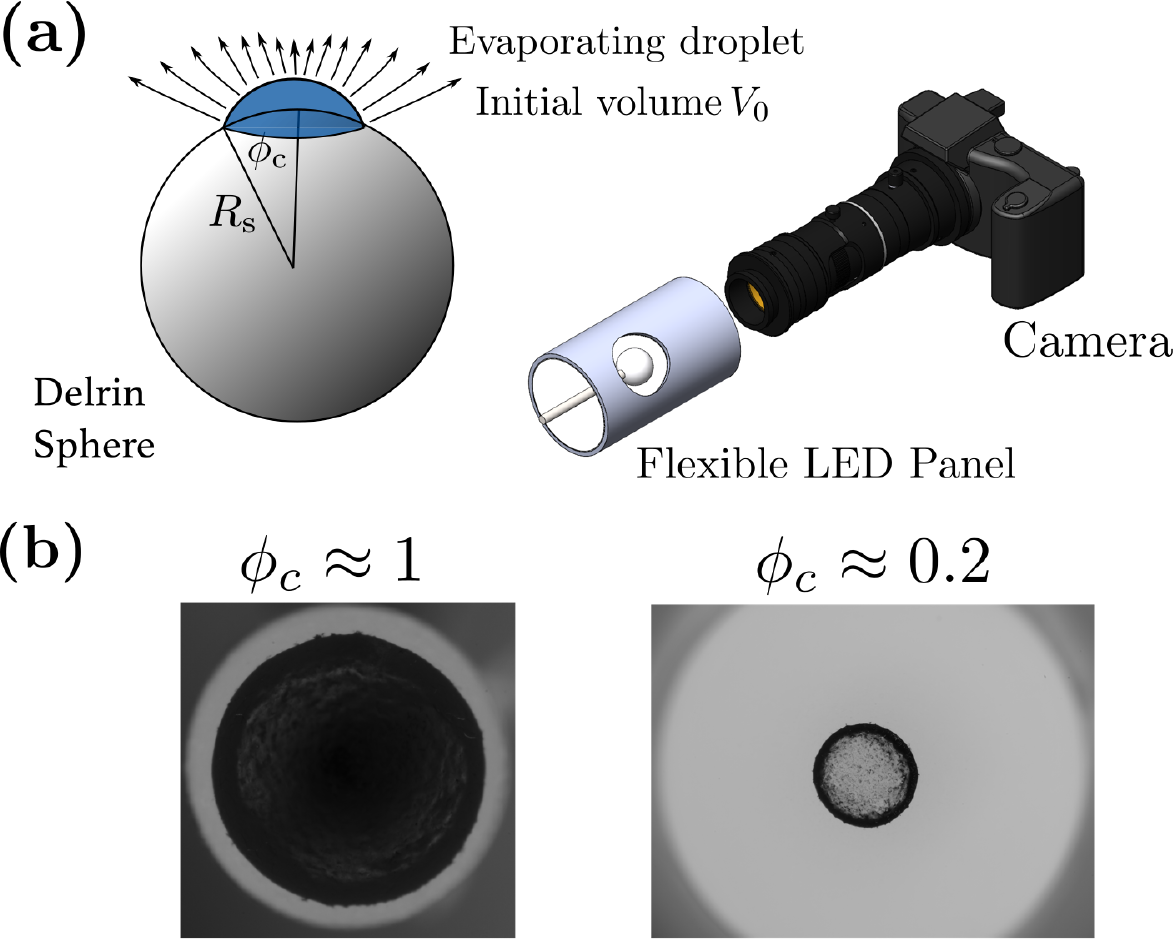}
\caption{(a) Experimental setup for studying evaporation stains formed on curved substrates. Schematic of a droplet on a sphere defining the angle $\phi_c$ (left) and showing how photos were captured once the droplet has evaporated leaving a stain behind (right). (b) Characteristic images from experiments with drops with a volume of 5$\mu$L on spheres of radii 1 mm (left) and 4mm (right) corresponding to two different values of $\phi_c$.} 
\label{fig:setup}
\end{figure}

As a first indication that curvature does have an effect consider Fig.\ 1(b): two droplets of the same volume are deposited on spheres of different radii, leading to different angular positions of the contact line. We observe that the stain on the smaller sphere (larger $\phi_c$) does not show a clear ring, while the stain on the larger sphere (smaller $\phi_c$) does. To characterize the transition from ring-like stains to more uniform stains we introduce the proportion of the droplet contact area covered by the stain, which was determined as follows: greyscale images were inverted and intensity profiles calculated so that high intensity corresponds to more prominent deposit. The ``covering" is then defined as the fraction of the intensity profile for which the intensity is above half of the maximum intensity.

\begin{figure}
    \centering
    \resizebox{.95\linewidth}{!}{\input{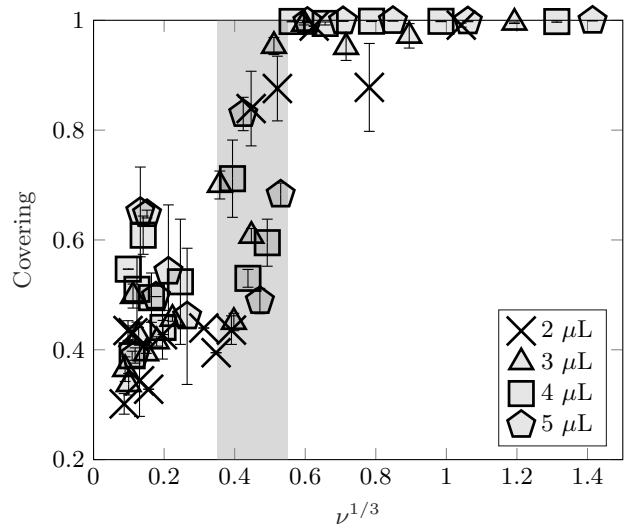}}
    \caption{The stain covering (defined as the fraction of the droplet footprint covered by deposit) as a function of the ratio of droplet to sphere volume, $\nu=\Vd/V_s$, which is our control parameter. Here $\nu^{1/3} \propto \phi_c$, the angular position of the contact line. As the curvature of the substrate --- relative to the droplet --- increases, more uniform stains form. The transition region (shaded grey area) is narrow and corresponds to a sharp transition; crucially, this transition region does not show systematic  variation as the droplet volume (indicated by symbol according to the inset legend) varies.}
    \label{fig2}
\end{figure}

For a given liquid-solid combination, the  initial droplet-sphere volume ratio is the only control parameter in our experiments. This ratio is denoted $\nu = \Vd/V_{s} \sim \phi_c^3$ assuming a constant contact angle at deposition \cite{SI}.  Figure \ref{fig2} shows the covering as a function of $\nu^{1/3} \sim \phi_c$: a sharp transition from ring-type stains to uniform stains occurs at $\nu^{1/3} \in [0.4,0.6]$; that this threshold appears independent of $\Vd$ suggests this is a geometrical effect, at least for low inital particle concentrations.

The criterion for the formation of a ring versus a uniform stain has been studied previously in the case of a sessile droplet on a planar substrate \cite{kaplan2015}. By comparing the ratio of two time scales  --- evaporation and advection --- it was found that the aspect ratio of the droplet and the capillary number of the flow determined the type of stain that would form. However, a similar scaling  argument applied to the case of a spherical substrate shows that increasing the control parameter $\nu$ in fact \emph{promotes} ring formation due to a non-vanishing gradient in Laplace pressure \cite{SI}. This is contrary to the experimental results presented in Fig.~\ref{fig2}. We now develop a more detailed model of the early stages of its formation by combining the modification to evaporation rate with flow on the sphere's surface.    

As already discussed, substrate curvature might be expected to affect droplet evaporation by modifying the evaporative flux but it may also modify the resulting flow within the droplet that leads to the CRE. To compare the importance (and direction) of these effects, we develop a model of a thin sessile suspension droplet on a spherical substrate of radius $R_s$, as shown in Fig.~\ref{fig:setup}(a). If we assume that the suspension is sufficiently dilute that particle interactions are negligible, the flow of solute can be described by simply considering conservation of mass. In the following, we first find the local evaporative flux from a droplet on a sphere, $J(\phi;\phi_c)$. Using this, we solve for the evaporation-induced flow inside the droplet, allowing us to find the particle concentration field inside the droplet while it remains dilute. This type of model --- relying solely on conservation of mass --- has been shown to offer insights in the planar case \cite{deegan2000}.   

Following \citet{popov2005} for the planar case, we assume that the evaporation of the droplet is limited by diffusion in the vapour and, hence, that the vapour concentration field is determined by the diffusion equation. Crucially, the time scale over which the vapour field changes, $t_{\mathrm{diff}}=R_s^2/D$, is much shorter than the lifetime of a droplet.  (For a 1 mm droplet with vapour diffusivity $D \sim 2.5 \times 10^{-5} \mathrm{ m}^2 / \mathrm{s}$ \cite{bergman2011}, $t_{\mathrm{diff}}\approx0.03\mathrm{~s}$) while the lifetime of the droplet is of the order of minutes.) We therefore assume that the concentration field of vapour, denoted $c_{\text{vap}}$ (mol /$\mathrm{m}^3$), is determined quasi-steadily and thus obeys Laplace's equation. 

Exploiting the spherical symmetry of the problem, we use axisymmetric spherical polar coordinates $r, \phi$ with the origin located at the centre of the sphere. We also assume the droplet's thickness to be small in comparison to the radius of the sphere, modelling it as a wet patch on the spherical substrate over a contact region  (as is common in planar evaporation models \cite{dunn2008, wray2020, dambrisio2021}); we denote the contact region $0<\phi<\phi_c$. At the interface of the droplet this concentration is assumed to be at saturation ($c_0$) and in the far-field ($r\rightarrow \infty$) $c_{\text{vap}}(r,\phi) \rightarrow c_{\infty}$. On the dry part of the sphere, there is no flux so we require $\partial_rc_{\text{vap}}(R_s,\phi) = 0$ for $\phi_c<\phi<\pi$. The problem for $c_{\text{vap}}$ can  be summarized as 
\begin{subequations}
\begin{align}
     &\nabla^2 c_{\text{vap}} = 0,\\
     &c_{\text{vap}}(r =R_s,\phi) = c_0, \quad 0 < \phi \leq \phi_c,\\
    &\partial_rc_{\text{vap}} (r = R_s, \phi) = 0, \quad \phi_c < \phi \leq \pi\\ 
    &c_{\text{vap}} \rightarrow c_{\infty}\quad \text{ as }\quad r \rightarrow \infty. 
\end{align}
\end{subequations}

We develop a semi-analytical approach to solve this problem \cite{SI}. This calculation results in an expression for the local evaporative  flux from the droplet surface, which may be written 
\begin{align}
    J(\phi;\phi_c) = \frac{DM}{\rho} \frac{\partial c_{\text{vap}}}{\partial r}\bigg|_{r=R_s},
    \label{eqn:EvapFlux}
\end{align}
where $D$ is the diffusion constant, $M$ is the molar mass and $\rho$ the liquid density. 
\begin{figure}
    \centering
    \resizebox{.95\linewidth}{!}{\input{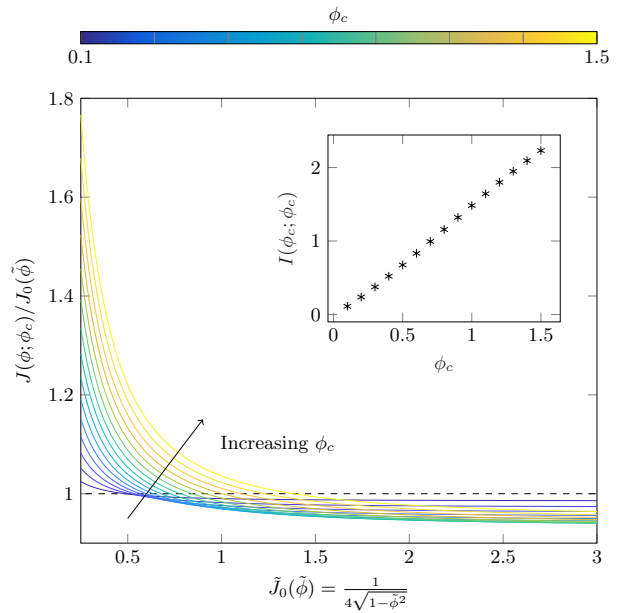}}
    \caption{The evaporative flux $J(\phi;\phi_c)$ compensated by that in the planar case, $J_0(\tphi)$, and plotted as a function of the dimensionless evaporative flux in the planar case ($\phi_c\rightarrow 0$).
As $\phi_c$ increases, the evaporative flux is enhanced at the centre of the drop while the singularity at the contact line becomes weaker. The dashed line corresponds to the planar case, $\phi_c = 0$. Inset: The integrated flux,  $I(\phi_c;\phi_c)$ defined in \eqref{eqn:DefnI} as computed numerically.}
    \label{fig:Flux}
\end{figure}

To understand how to compare the evaporative flux at different values of $\phi_c$ we first consider the planar limit, which is recovered for fixed contact line radius by the limit $\phi_c \rightarrow 0$, $R_s\to\infty$ with the product $R_s\phi_c$ fixed. In this limit, a detailed analysis (see \cite{SI}) shows that the  evaporative flux has leading order behaviour
\begin{equation}
J(\phi;\phi_c) \approx J_0(\tphi) = \frac{DM(c_0-c_\infty)}{\rho\, R_s\phi_c}\frac{2}{\pi}(1-\tphi^2)^{-1/2}
\label{eqn:EvapFlux0}
\end{equation}
where $\tphi = \phi/\phi_c$ is a rescaled angular coordinate. This reproduces precisely the planar limit for a droplet with contact line radius $r^{2\mathrm{D}}_c=R_s\phi_c$ and rescaled radial coordinate $\tilde{\phi}\to r^{2\mathrm{D}}/r^{2\mathrm{D}}_c$ \cite{popov2005}, where $r^{2\mathrm{D}}$ denotes the radial position for cylindrical polar coordinates centered with the drop.

Figure \ref{fig:Flux} shows a plot of the evaporative flux compensated to facilitate comparison with the planar case. This plot shows that as $\phi_c$ increases, the evaporative flux at the centre of the droplet increases while the flux at the contact line decreases --- as might be expected from the molecular view of evaporation, redistributing neighbouring liquid molecules changes the distribution of evaporation across the droplet.

The evaporative flux from \eqref{eqn:EvapFlux} induces a flow in the fluid. To calculate this flow, we note that the capillary pressure dominates viscous stress: the relevant capillary number $\mathrm{Ca}=\mu u/(2\gamma \theta) \approx 10^{-7}\ll1$ \cite{Marin2011} (where $\theta$ denotes contact angle).  Hence the droplet's interface, located at $r=R_s+h(\phi,t)$, is determined quasi-statically and is simply a spherical cap (with centre displaced from the sphere's centre in general).  The flow within the thin droplet is then determined solely by conservation of mass, which can be written
\begin{align}
\frac{\partial h}{\partial t} + \frac{1}{R_s \sin\phi}\frac{\partial}{\partial \phi} \left( \sin\phi h \bar{u} \right) = J(\phi;\phi_c),\label{cof}
\end{align}
with $\bar{u}$ the depth-averaged fluid velocity. 

We assume that the particles are well mixed throughout the thickness of the  droplet (so that the solute concentration $c(\phi,r,t) = \bar{c}_{\text{sol}}(\phi,t)$ only) and that they are simply advected with the flow towards the contact line, neglecting particle diffusion and particle interactions such as jamming. Thus, conservation of solute can be written as 
\begin{equation}
    \frac{\partial \bar{n}}{\partial t} + \frac{1}{R_s\sin\phi} \frac{\partial}{\partial \phi} \left( \bar{u} \bar{n} \sin\phi \right) = 0, \label{cos}
\end{equation}
where $\bar{n}(\phi,t) = h(\phi,t)\bar{c}_{\text{sol}}(\phi,t)$ is the mass density (per unit projected surface area) of particles at a particular point. 

The quantity of most interest in this study is the amount of solute mass at a particular position, $\bar{n}(\phi,t)$. This is determined by the solution of \eqref{cof}--\eqref{cos}, with $J(\phi;\phi_c)$ given by \eqref{eqn:EvapFlux}. To facilitate this solution, we use the droplet lifetime
\begin{equation}
    t_f = \Vd / [4\alpha R_s I(\phi_c;\phi_c)],
    \label{eqn:tf}
\end{equation} to rescale time, where $\alpha = DM(c_0 - c_{\infty})/\rho$; this also leads to a characteristic velocity scale $u_s = R_s \phi_c / t_f$. Here
\begin{equation}
I(\phi_c;\phi_c) = -\frac{\pi R_s}{2 \alpha}\int_{0}^{\phi_c} J(\phi;\phi_c) \sin(\phi) \, \mathrm{d}\phi. 
\label{eqn:DefnI}
\end{equation}
is the integrated evaporation flux, and is plotted in the inset of fig.~\ref{fig:Flux}. To solve this problem, we supplement it by initial conditions such that the initial profile of the drop corresponds to a volume $\Vd$ and uniform solute concentration $c_{\mathrm{sol}}(\phi,0)=C_0$.

\begin{figure}
       \centering
       \resizebox{.95\linewidth}{!}{\input{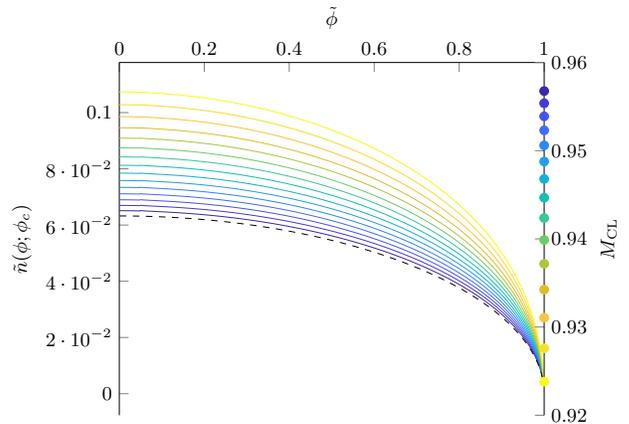}}
       \caption{The dimensionless solute area density, $\tilde{n}$ evaluated at $\tau = 0.99$ for $\phi_c$ ranging from 0.1 - 1.5. The dashed line denotes the  planar case, $\phi_c=0$. The value of $\tilde{n}$ within the droplet is shown by different curves while the solute mass accumulated at  the contact line, $m_{\text{CL}}$, is shown by points on the right-hand axis. As substrate curvature increases ($\phi_c$ increases), the amount of solute remaining in the interior of the droplet after a fixed fraction of the evaporation increases by a factor of up to 2.} 
       \label{fig:conc}
  \end{figure}

Since the drop profile $h(\phi,t)$ is determined quasistatically by the droplet volume, Eqn \eqref{cof} can readily be solved numerically to give the flow field $\bar{u}$. With $\bar{u}$ determined, \eqref{cos} can be solved numerically using the method of characteristics to give  $\bar{n}(\phi,t)$, the instantaneous mass of solute contained, per unit surface area of the sphere, at angular position $\phi$ --- this is the quantity of primary interest in determining  the final stain. However, a simple comparison of $\bar{n}$ with different $\phi_c$ appears confounded since the total mass of solute, $m=2\pi R_s^2 \int_0^{\phi_c}\bar{n}\sin\phi~d\phi$, is \emph{not} conserved:  the advection of fluid carries solute to the contact line where it is lost to the system. We account for this by introducing the solute mass at the contact line, $m_{\mathrm{CL}}$, to ensure that the mass of solute is globally conserved, i.e.
\begin{equation}
m_{\mathrm{CL}}(t;\phi_c)= C_0 \Vd - 2\pi R_s^2 \int_0^{\phi_c} \bar{n}(\phi,t; \phi_{c}) \sin\phi \,\mathrm{d}\phi
\end{equation}
We scale the solute density and mass at the contact line by letting 
\begin{eqnarray*}
    \bar{n}(\phi,t;\phi_c) &= \frac{c_0\Vd\, }{2 \pi R_s^2 (1 - \cos \phi_c)} \tilde{n}(\phi,t;\phi_c)\\ m_{\mathrm{CL}}(t;\phi_c)&=C_0\Vd M_{\mathrm{CL}}(t;\phi_c). 
\end{eqnarray*}
A plot of $\tilde{n}(\phi,0.99t_f;\phi_c)$ is shown in Fig.~\ref{fig:conc} for a range of $\phi_c$.  It shows that as $\phi_c$ increases more mass is retained within the bulk of the droplet --- particularly towards the centre. Conversely the amount that has traveled to the contact line, $m_{\text{CL}}$ decreases as $\phi_c$ increases (see the right-hand axis of Fig.~\ref{fig:conc} for $m_{\mathrm{CL}}(0.99t_f;\phi_c)$). 

Our model shows behaviour qualitatively similar to experiments but does not immediately highlight the relative importance of changes to the flow  and evaporative flux that are caused by geometry. We therefore consider asymptotically the model behavior for $\phi_c\ll1$ (see SI) and find that perturbations to the evaporative flux enter at $O(\phi_c)$, while corrections to the droplet profile $\tilde{\eta}$ only enter at order $O(\phi_c^2)$. Consequently, there is an $O(\phi_c)$ correction to the flow field $\bar{u}$ caused by the geometry-induced modification to the evaporative flux; we conclude that suppression of the CRE by curvature is associated with the weakening of the evaporative singularity at the contact line and enhanced evaporation at the centre of the droplet. 

This simple model shows that substrate curvature can lead droplets to have larger concentration of solute throughout the droplet. In our model, all particles are eventually advected to the contact line, but we note that other works have shown that higher concentration slows the evaporation rate and promotes the formation of uniform stains \cite{kim2018}. We do not consider precisely how particle interactions such as jamming \cite{popov2005} or increasing viscosity \cite{kaplan2015} may affect the final stain, merely suggesting that consideration of these effects may help to explain the rather sharp transition from rings to uniform stains present in Fig.~\ref{fig2}.

We have shown that substrate curvature can modify the evaporative flux of a droplet, which in turn leads to more solute remaining in the bulk of the droplet (rather than at the contact line), thereby increasing the uniformity of the stain produced. This effect might lend itself to being utilized in many industrial processes that rely on inkjet printing, including flexible electronics. For example, if the substrate is a thin elastic material, one could imagine temporarily increasing the curvature (for example by inflation),  depositing a solute droplet on the (now curved) substrate and allowing it to dry to leave a uniform stain. Upon deflation the substrate would return to its flat configuration, as would the  stain. Realizing this in practice would require complications, such as the elastic strain induced in the uniform stain by deflation, to be accounted for and hence is left as a topic for future work.

This research was funded by the Royal Society (URF\textbackslash{}R\textbackslash{}180016, and RGF\textbackslash{}EA\textbackslash{}181002) (AACP), and supported by the Leverhulme Trust (DV). We would also like to thank J.~Oliver, M.~Moore and T.~du Fayet de la Tour for discussions in the early stages of this work.

\newpage
\appendix

\bibliography{apssamp}

\newpage
\onecolumngrid
\renewcommand{\thefigure}{S\arabic{figure}}
\setcounter{figure}{0}

\section*{Supplementary Material}

In this Supplementary Material, we apply the scaling argument of Kaplan \& Mahadevan \cite{kaplan2015} to the spherical problem of interest here (Appendix \ref{sec:scaling}) and provide further details of our experimental protocols (Appendix \ref{sec:expts}). The remainder of the Supplementary Material (Appendix \ref{sec:CalcDetails}) gives further details of the model calculations used to produce figures 3 and 4 of the main text.

\section{Scaling Argument \label{sec:scaling}}
In this section, we apply the scaling argument of \citet{kaplan2015} as a first attempt to explain the observed effect of substrate curvature on the coffee ring effect. 

\begin{figure}
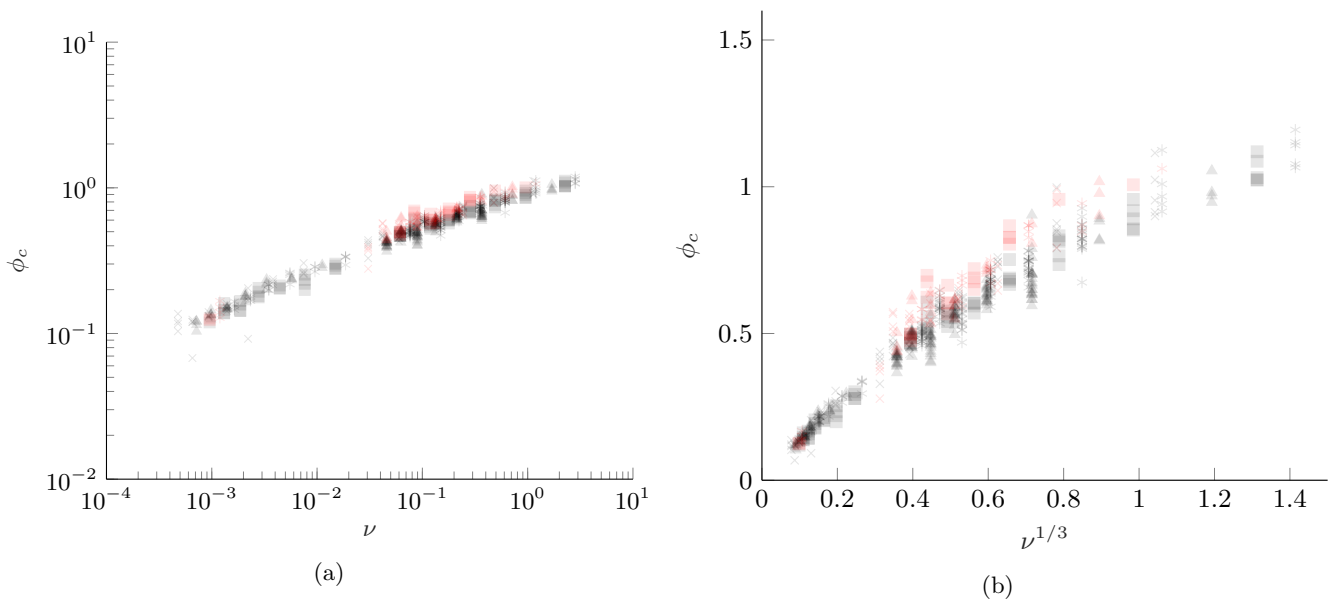

\begin{subfigure}{0.49\linewidth}
\resizebox{\textwidth}{!}{\input{loglofplotnupc}}
\caption{}
\end{subfigure}
\centering
\begin{subfigure}{0.49\linewidth}
\resizebox{\textwidth}{!}{\input{linearplotnupc}}
\caption{}
\end{subfigure}
\caption{Equation \eqref{eqB5} suggests that our experimental parameter $\nu \sim \phi_c^3$ in the small $\phi_c$ limit. Experimental results, points, show that this scaling holds for $\phi_c$ up to $\phi_c\approx 0.8$. }
\label{scaling}
\end{figure}

The approach of \citet{kaplan2015} relies on comparing the advection timescale with that induced by evaporation. This ratio, $\alpha = t_{\mathrm{adv}}/t_{\mathrm{evap}}$ gives a sense of whether particles have time to be advected to the contact line or are simply deposited in their initial location. 
To get a scaling for the fluid velocity we appeal to the lubrication equation in spherical coordinates, i.e.
\begin{equation}
    \mu \frac{\partial^2 u}{\partial r^2} = \frac{1}{r}\frac{\partial P}{\partial \phi}. 
\end{equation}
In scaling terms then 
\begin{equation}
    \frac{\mu u_f}{H^2} \sim \frac{\Delta P}{R_s \phi_c}
\end{equation}
and hence the fluid velocity is
\begin{equation}
    u_f \sim \frac{\Delta P H^2}{R_s \phi_c \mu}. 
\end{equation}
The volume of a droplet on a sphere is given as, 
\begin{eqnarray}
\Vd = \frac{\pi}{3}R_d^3 T(\phi_c + \theta) - \frac{\pi}{3}R_s^3 T(\phi_c),
\end{eqnarray}
where $T(x) = (2+\cos x) (1 - \cos x)^2$, $R_d$ is the radius of curvature of the droplet. For small angles this shows that 
\begin{eqnarray}
\Vd \sim (R_s \phi_c )^3 \theta \label{eqB5}.
\end{eqnarray} As a result, we expect the experimental control parameter $\nu=\Vd/V_s\propto \phi_c^3$, as is confirmed experimentally (see Fig.~\ref{scaling}).

The quantities $H$ and $\Delta P$ can be estimated from geometry as, 
\begin{align}
    H &\sim {R_s \phi_c \theta},\\
    \Delta P &\sim \gamma / R_d \sim \frac{\gamma}{R_s \phi_c}(\phi_c + \theta),
\end{align}
where $R_d$ denotes the radius of curvature of the droplet and $\theta$ is the contact angle of the droplet. Thus,
\begin{equation}
    u_f \sim \frac{\gamma \theta^2}{ \mu}(\phi_c + \theta). 
\end{equation}
The timescale for advection then is 
\begin{equation}
    t_{\mathrm{adv}} = \frac{R_s \phi_c}{u_f} = \frac{\mu R_s \phi_c}{\gamma {\theta^2(\phi_c + \theta)} },
\end{equation}
while the timescale for evaporation is
\begin{equation}
    t_{\mathrm{evap}} = H/E 
\end{equation}
where $E$ denotes the evaporation rate. 

The parameter $\alpha$ is 
\begin{equation}
\alpha = \frac{t_\text{evap}}{t_\text{adv}} = \frac{H}{E}\frac{u_f}{R_s\phi_c} =  \frac{\gamma \theta^3}{\mu E}(\phi_c  +\theta).
\end{equation}
According to the intuition from \citet{kaplan2015}, $\alpha$ increasing corresponds to more time for particles to be advected to the contact line, hence the stain is expected to be more ring-like. However, this scaling argument shows that $\alpha$ increases with $\phi_c$ so that sphere curvature makes the formation of rings (rather than a uniform stain) more likely --- precisely the opposite of our experimental findings. This is because the above argument does not account for the change in evaporation induced by substrate curvature.

\section{Experimental Details\label{sec:expts}}
\begin{figure}
    \centering
    \resizebox{.45\linewidth}{!}{\input{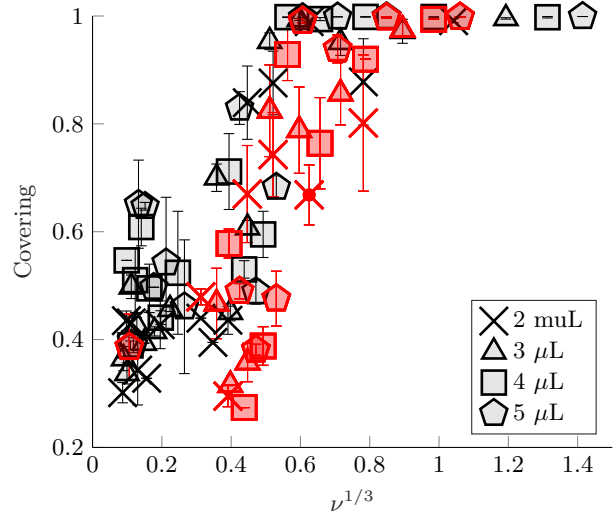}}
    \caption{The covering (fraction of droplet footprint covered) as a function of $\left( \Vd/V_s \right)^{1/3} = \nu^{1/3} \propto \phi_c$, which is the experimental control parameter. Results are shown for a variety of droplet volumes (see legend) and in different orientations:  black markers denote droplets in the pendant configuration while red markers denote droplets in the sessile configuration.}
    \label{fig2app}
\end{figure}

\begin{figure}
    \centering
    \resizebox{.45\linewidth}{!}{\input{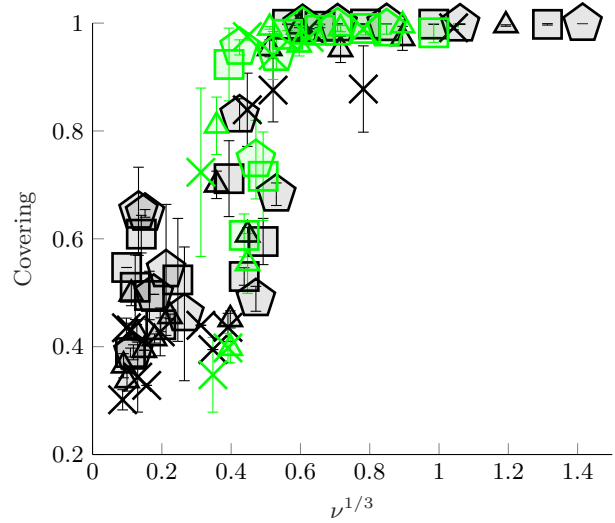}}
    \caption{The covering (fraction of droplet footprint covered) as a function of $\left( \Vd/V_s \right)^{1/3} = \nu^{1/3} \propto \phi_c$, which is the experimental control parameter. Results are shown for a variety of droplet volumes (see legend of fig.~\ref{fig2app}) and initial particle concentrations:  black markers denote droplets with an initial concentration of 0.1\% while green markers denote droplets of initial concentration of 0.15\%. In both sets of experiments,  droplets are in the pendant configuration.}
    \label{fig3app}
\end{figure}

Our experiments consisted of placing droplets of nigrosine solution on Delrin spheres (purchased from \texttt{www.ballandrollerstore.com}). The spheres were attached to slender needles using hot glue. The evaporating droplets were kept a distance of at least $\sim$ 7 cm apart to avoid effects such as shielding \cite{wray2021}. To rule out the effect of gravity on our observations, we evaporated the droplets both on the sphere as shown in Fig.\ 1(a) --- the sessile configuration --- but also hanging upside down from the sphere --- the pendant configuration. The covering as a function of $\phi_c$ in each configuration can be seen in Fig.~\ref{fig2app}; we conclude from this that the effect of the orientation (sessile \emph{versus} pendant) is minimal. The effect of initial concentration is shown in Fig.~\ref{fig3app} where results for two solution concentrations are shown.

Once the droplets had evaporated, photos were taken of the resulting stain as in Fig.\ 1(b). From these photos, the area of the stain could be extracted and an average value of $\phi_c$ calculated. We find that the control parameter $\nu$ scales as $\phi_c^3$, see fig.~\ref{scaling} which can be explained from Equation \eqref{eqB5}.

\section{Calculation Details\label{sec:CalcDetails}}

\subsection{Solving Laplace's Equation}
In the following we present our solution for the concentration field $C$. We first transform the full problem (Equations 1(a)-(d)) to one with simpler boundary conditions by letting:
\begin{equation}
c_{\text{vap}} = (c_0 - c_{\infty})C + c_{\infty}.
\end{equation}

The problem is then to solve Laplace's equation
\begin{equation}
\nabla^2C=0\;,
\label{eq:Laplace}
\end{equation}
subject to the boundary conditions
\begin{eqnarray}
&&C(r\rightarrow\infty,\phi)\rightarrow 0\;,\label{eq:bc1}\\
&&C(r=R_s,\phi)=1\quad 0\leq\phi\leq\phi_c\;,\label{eq:bc2}\\
&&\frac{\partial C}{\partial r}(r=R_s,\phi) =0 \quad \phi_c<\phi\leq\pi\;.\label{eq:bc3}
\end{eqnarray}

To exploit the symmetry of the problem, we use spherical coordinates $(r,\phi,\theta)$ with $0\leq\phi\leq\pi$, $0\leq\theta\leq 2\pi$ and $r\geq R_s$ (corresponding to the region beyond the sphere). Assuming axisymmetry, we ignore variation in the  $\theta$ direction so that $C = C(r,\phi)$. 

 
The axisymmetric solution of Eq.~(\ref{eq:Laplace}) is given by
\begin{equation}
C(r,\phi)= \sum_{n=0}^{\infty}  \frac{A_n}{(n+1)} \left(\frac{R_s}{r}\right)^{n+1}P_n(\cos \phi)\;,
\end{equation}
where $P_n$ is the Legendre Polynomial and $A_n$ are real constants. The contribution coming from $Q_n(\cos\phi)$ vanishes because it has a singular behavior at $\phi=0$ or $\phi=\pi$ while the far-field behaviour is guaranteed by the choice of $r$-dependence. 

Written in this way, the boundary conditions~(\ref{eq:bc2},\ref{eq:bc3}) become
\begin{eqnarray}
&&\sum_{n=0}^{\infty}  \frac{A_n}{(n+1)} P_n(\cos \phi)=1\qquad 0<\phi<\phi_c\;, \label{eqn:Ser1}\\
&&\sum_{n=0}^{\infty} A_n P_n(\cos \phi)=0\qquad \phi_c<\phi<\pi\;\label{eqn:Ser2}.
\end{eqnarray}
To solve Equations \eqref{eqn:Ser1}--\eqref{eqn:Ser2} simultaneously we reduce the problem to an integral equation. 

Using well known properties of Legendre Polynomials \cite{gradshteyn2014} we have that,
\begin{equation}
\sum_{n=0}^\infty \cos\left[(n+1/2)u\right] P_n(\cos\phi)=\frac{H(\phi-u)}{\sqrt{2(\cos u-\cos\phi)}}\; \label{eqn:LPprop1},
\end{equation}
and
\begin{equation}
\sum_{n=0}^\infty \sin\left[(n+1/2)u\right] P_n(\cos\phi)=\frac{H(u-\phi)}{\sqrt{2(\cos\phi-\cos u)}}\; \label{eqn:LPprop2}.
\end{equation}
for all $0<\phi<\pi$ and $0<u<\pi$. Here $H(\cdot)$ is the Heavside function.

Let us suppose that 
\begin{equation}
A_n=\frac{2}{\pi}(n+1/2)\int_0^{\phi_c}  g(u)\cos \left[(n+1/2)u\right]\mathrm{d}u\;,
\label{eqn:CoeffsG}
\end{equation}
for some $g(u)$. Integrating by parts, we have:  
\begin{equation}
A_n=\frac{2}{\pi}g(\phi_c)\sin \left[(n+1/2)\phi_c\right]-\frac{2}{\pi}\int_0^{\phi_c}  g'(u)\sin \left[(n+1/2)u\right]\mathrm{d}u\; \label{eqn:AnIBP},
\end{equation}
provided that $g(\phi)$ is regular at $\phi=0$. 
The $r$-derivative of $C$ evaluated at $r=R_s$ is given by, 
\begin{eqnarray}
\frac{\partial C}{\partial r}(r=R_s,\phi)= -\frac{1}{R_s} \sum_{n=0}^\infty A_n P_n(\cos\phi). 
\end{eqnarray}
Using Equation \eqref{eqn:AnIBP} we can write this derivative as,
\begin{eqnarray}
\frac{\partial C}{\partial r}(r=R_s,\phi) = - \frac{2 g(\phi_c)}{R_s \pi} \sum_{n=0}^\infty \sin[(n+1/2)\phi_c]P_n(\cos\phi) + \frac{2}{R_s \pi} \sum_{n=0}^\infty P_n(\cos\phi) \int_{0}^{\phi_c} g'(u) \sin[(n+1/2)u] \mathrm{d}u
\end{eqnarray}
The properties of Legendre Polynomials \eqref{eqn:LPprop1} and \eqref{eqn:LPprop2} then give
\begin{equation}
 -\frac{\partial C}{\partial r}(r=R_s,\phi)=\frac{2}{\pi R_s}H(\phi_c-\phi)\left[\frac{g(\phi_c)}{\sqrt{2(\cos\phi-\cos \phi_c)}}-\int_\phi^{\phi_c}  \frac{g'(u)\mathrm{d}u}{\sqrt{2(\cos\phi-\cos u)}}\right]\;,
 \label{eqn:fluxlocal}
\end{equation}
which can be rewritten as
\begin{equation}
\frac{\partial C}{\partial r}(r=R_s,\phi)=\frac{2}{\pi R_s}H(\phi_c-\phi)\frac{1}{\sin\phi}\frac{\mathrm{d}}{\mathrm{d}\phi}\int_\phi^{\phi_c}  \frac{g(u)\sin u \,\mathrm{d}u}{\sqrt{2(\cos\phi-\cos u)}}\;.
\end{equation}
This solution automatically satisfies the boundary condition~(\ref{eq:bc3}) and also shows that, in general, the flux will be singular as $\phi \rightarrow \phi_c$ i.e.
\begin{eqnarray}
\frac{\partial C}{\partial r}(r=R_s,\phi) \propto \frac{H(\phi_c - \phi)}{\sqrt{\phi_c - \phi }} \quad \phi \rightarrow \phi_c.
\end{eqnarray}
The flux can be computed once $g(u)$ is known.

Now, replacing $A_n$ in the boundary condition~(\ref{eq:bc2}) results in the following integral equation
\begin{equation}
\int_0^{\phi}   \frac{g(u)\mathrm{d}u}{\sqrt{2(\cos u-\cos \phi)}}=\frac{\pi}{2}+\int_0^{\phi_c}  K(u,\phi) g(u)\,\mathrm{d}u\;,
\label{eq:integral}
\end{equation}
for all $0<\phi<\phi_c$. Here,
\begin{equation}
K(u,\phi)=\frac{1}{2}\sum_{n=0}^{\infty} \frac{\cos\left[(n+1/2)u\right] P_n(\cos \phi)}{(n+1)}\;.
\end{equation}
The integral equation~(\ref{eq:integral}) can be inverted using an Abel transform. First we rewrite it as
\begin{equation}
\int_0^{\phi}   \frac{g(u)\mathrm{d}u}{2\sqrt{\sin^2[\phi/2] -\sin^2[u/2]}}=\frac{\pi}{2} +\int_0^{\phi_c}  K(u,\phi) g(u)\,\mathrm{d}u\;.
\end{equation}
The inversion yields
\begin{equation}
\frac{g(\phi)}{\cos[\phi/2]}=\frac{2}{\pi}\frac{\mathrm{d}}{\mathrm{d}\sin[\phi/2]}\left[\int_0^{\sin[\phi/2]}\frac{\rho \mathrm{d}\rho}{\sqrt{\sin^2[\phi/2]-\rho^2}} \,\left(\frac{\pi}{2} +\int_0^{\phi_c}  K(u,\rho) g(u)\,\mathrm{d}u\right)\right]\;,
\end{equation}
\begin{equation}
\frac{g(\phi)}{\cos[\phi/2]}=1+\frac{2}{\pi}\int_0^{\phi_c}  \mathrm{d}u\, g(u)\left[\frac{\mathrm{d}}{\mathrm{d}\sin[\phi/2]}\int_0^{\sin[\phi/2]}\frac{\rho K(u,\rho) \mathrm{d}\rho}{\sqrt{\sin^2[\phi/2]-\rho^2}}\right]\;.
\end{equation}
Thus the result of the inversion gives
\begin{equation}
g(\phi)=\cos[\phi/2]+\int_0^{\phi_c} Q(u,\phi) g(u)\,\mathrm{d}u\;,
\end{equation}
with
\begin{equation}
Q(u,\phi)=\frac{1}{\pi}\sum_{n=0}^{\infty} \frac{\cos\left[(n+1/2)u\right] }{(n+1)}\frac{\mathrm{d}}{\mathrm{d}\phi}\int_0^{\phi}\frac{ P_n(\cos t) \sin t\, \mathrm{d} t}{\sqrt{2(\cos t-\cos \phi)}}\;.
\end{equation}

The kernel $Q$ can be computed analytically since,
\begin{equation}
\int_x^1\frac{P_n(t)\,\mathrm{d}t}{\sqrt{t-x}}=\left(n+1/2\right)^{-1}\left(1-x\right)^{-1/2}(T_n(x)-T_{n+1}(x))\;,
\end{equation}
as found in \citet{gradshteyn2014}. We therefore have
\begin{align}
Q(u,\phi) &=\frac{\sqrt{2}}{\pi}\sum_{n=0}^{\infty} \frac{\cos\left[(n+1/2)u\right] }{(n+1)(2n+1)}\frac{d}{d\phi}\left[\frac{ T_n(\cos \phi) -T_{n+1}(\cos\phi)}{\sqrt{1-\cos\phi}}\right]\;,\\
&=\frac{\sqrt{2}}{\pi}\sum_{n=0}^{\infty} \frac{\cos\left[(n+1/2)u\right] }{(n+1)(2n+1)}\frac{d}{d\phi}\left[\frac{ \cos n\phi -\cos(n+1)\phi}{\sqrt{1-\cos\phi}}\right]\;,\\
&=\frac{1}{2\pi}\sum_{n=0}^{\infty} \frac{\cos\left[(n+1/2)(\phi+u)\right]+ \cos\left[(n+1/2)(\phi-u)\right]}{(n+1)}\;.
\end{align}
One can thus write 
\begin{equation}
2\pi\,Q(u,\phi)=I(\phi+u)+ I(\phi-u)\;,
\end{equation}
with
\begin{equation}
I(x)=\sum_{n=0}^{\infty} \frac{\cos(n+1/2)x}{(n+1)}\;.
\end{equation}
Computer algebra gives:
\begin{equation}
I(x)=-\frac{1}{2}   \left(e^{i x/2}\log\left(1 - e^{-i x}\right) + e^{-i x/2} \log\left(1 - e^{i x}\right)\right)\;.
\end{equation}
which can be simplified to give,
\begin{equation}
I(x)= \frac{1}{2}(\pi-|x|)\sin[|x|/2]-\cos[ x/2]\log\left(2\sin[|x|/2]\right)\;.
\end{equation}
The integral equation for $g(\phi)$ is then given by
\begin{equation}
g(\phi)=\cos(\phi/2)+\frac{1}{2\pi}\int_0^{\phi_c} (I(\phi+u)+ I(\phi-u)) g(u)\,\mathrm{d}u\;.
\end{equation}

Since we have an analytical expression for $I(x)$, the numerical problem to compute $g(u)$ becomes simple. The integral equation can be solved using an iterative method (Picard's method). We assume
\begin{equation}
g(\phi)=\sum_{i=0}^N g_i(\phi)\;,
\end{equation}
with
\begin{equation}
g_0(\phi)=\cos(\phi/2)\;,
\end{equation}
and for $i>0$
\begin{equation}
g_{i}(\phi)=\frac{1}{2\pi}\int_0^{\phi_c} \left[I(\phi+u)+ I(\phi-u)\right] g_{i-1}(u)\,\mathrm{d}u\;.
\end{equation}
We terminate the iteration when $|g_N| < 10^{-11}$. The integrals are easily computed numerically since no harmful singular behavior is present. A little caution is required because of  the (weak and integrable) logarithmic singularity of $\log\left(\sin[x/2]\right)$ as $x\to0$; this can easily be handled using standard methods as described by, for example, \citet{acton1990}.   

\subsection{Evaporative flux and droplet lifetime}

The above analysis gives the concentration gradient to be: 
\begin{equation}
    \frac{\partial c}{\partial r} \bigg|_{r=R_s} = -\frac{2}{\pi} \frac{c_0 - c_\infty}{R_s} F(\phi;\phi_c),
\end{equation} where
\begin{equation}
    F(\phi;\phi_c) = \frac{g(\phi_c)}{\sqrt{2(\cos\phi - \cos\phi_c)}}    - \int_{\phi}^{\phi_c} \frac{g'(u)\, \mathrm{d}u}{\sqrt{2(\cos\phi - \cos u)}}.     
\end{equation}

The  rate at which the volume of the droplet, $V$, decreases due to evaporation is thus given by 
\begin{equation}
    \dot{V} = \int -\frac{2}{\pi} \frac{\alpha}{R_s} F(\phi;\phi_c) \, \mathrm{d}A
\end{equation} where $\alpha= MD(c_0-c_\infty)/\rho$. We thus have
\begin{equation}
      \dot{V} = -4\alpha R_s I(\phi_c;\phi_c),
\end{equation} where
\begin{equation}
      I(\phi;\phi_c) = \int_{0}^{\phi} F(\bar{\phi};\phi_c) \sin\bar{\phi} \, \mathrm{d}\bar{\phi}.
      \label{eqn:PhiDefn}
\end{equation}
We write this as
\begin{align}
    V(t) = \Vd(1-t/t_f),
    \end{align}
    where the lifetime of the droplet is
    \begin{align}
     t_f = \frac{\Vd}{4\alpha R_s I(\phi_c;\phi_c)}.
\end{align}

\subsection{Evaporation-induced fluid flow}

Conservation of fluid can be written as 
\begin{equation}
    \frac{\partial h}{\partial t} + \frac{1}{R_s \sin\phi} \frac{\partial}{\partial \phi} \left( \sin\phi h\bar{u} \right) = -\frac{2}{\pi}\frac{\alpha}{R_s}F(\phi;\phi_c),  
    \label{eqn:CoF_raw}
\end{equation}
where $h(\phi,t)$ is the droplet shape which we assume to be determined quasistatically. In particular, the shape of the droplet is given by, 
\begin{equation}
    h(\phi,t) = \frac{V(t)}{\pi R_s^2} \frac{\cos\phi - \cos\phi_c}{(1-\cos\phi_c)^2}.
\end{equation}
However, it is useful to write this as 
\begin{equation}
    h(\phi,t) = \frac{2V(t)}{\pi R_s^2 \phi_c^2} \tilde{\eta}(\phi;\phi_c),
\end{equation} where
    \begin{equation}
    \tilde{\eta}(\phi;\phi_c) = \frac{\phi_c^2}{2} \frac{\cos\phi - \cos\phi_c}{(1-\cos\phi_c)^2}\label{etatw} 
    \end{equation} is introduced as the `shape function' because, in small $\phi_c$ limit, $\tilde{\eta} \sim 1-\tphi^2$ --- this matches the spherical cap shape in the planar case, making direct comparison easier. 

From \eqref{eqn:CoF_raw}, the typical fluid velocity is $u_s = R_s \phi_c / t_f$, and so we let $\bar{u}=u_s\tilde{u}$; the dimensionless equivalent of \eqref{eqn:CoF_raw} is thus
\begin{equation}
    - \tilde{\eta} + (1-\tau) \frac{\phi_c}{\sin\phi}\frac{\partial}{\partial \phi} \left( \tilde{\eta} \tilde{u} \sin \phi   \right) = -\frac{ \phi_c^2}{4 I(\phi_c;\phi_c)} F(\phi;\phi_c),\label{eqn:COFapp}
\end{equation} where $\tau=t/t_f$ is the dimensionless time.

Equation \eqref{eqn:COFapp} can be integrated once to find
\begin{align}
   \tilde{u} &= \frac{w(\phi;\phi_c)}{1-\tau},
\end{align}
where
\begin{align}
    w(\phi;\phi_c) &= \frac{1}{\tilde{\eta} \sin\phi} \frac{\phi_c}{4} \left[1 -  \frac{(\cos\phi - \cos\phi_c)^2}{(1-\cos\phi_c)^2} - \frac{I(\phi;\phi_c)}{I(\phi_c,\phi_c)} \right]
\end{align} and $I(\phi;\phi_c)$ is defined in \eqref{eqn:PhiDefn}.

\subsection{Flow-induced evolution of the concentration field}

The dimensionless solute problem (Equation 5 of the main text) is
\begin{equation}
    \frac{\partial \tilde{n}}{\partial \tau} + \frac{\phi_c}{(1-\tau)\sin\phi} \frac{\partial}{\partial \phi} \left[ w(\phi;\phi_c)\tilde{n} \sin\phi \right] = 0. \label{eqn:solProb}
\end{equation}
It is helpful to rewrite this by letting
\begin{align}
    M &= w(\phi;\phi_c)\tilde{n}\sin\phi,
    \end{align}
    so that Equation \eqref{eqn:solProb} is rewritten
    \begin{align}
\frac{\partial M}{\partial\tau} &+ \frac{\phi_c w(\phi;\phi_c)}{1-\tau} \frac{\partial M}{\partial \phi} = 0.
\end{align}
This can then be solved using the method of characteristics: the value of $M$ is preserved along curves with $d \phi/d\tau=\phi_cw(\phi;\phi_c)/(1-\tau)$. We can therefore write that the concentration field $\tilde{c}$ is given by 
\begin{equation}
    \tilde{c}(\phi,\tau;\phi_c) = \frac{1}{1-\tau}\left[\frac{\sin\phi_0}{\sin\phi}\frac{\tilde{\eta}(\phi_0;\phi_c)}{\tilde{\eta}(\phi;\phi_c)}\frac{w(\phi_0;\phi_c)}{w(\phi;\phi_c)}\right] 
\end{equation}
where $\phi_0$ is the solution to,
\begin{equation}
    -\phi_c \log(1-\tau) = \int_{\phi_0}^{\phi} \frac{1}{w(\bar{\phi},\phi_c)} \, \mathrm{d}\bar{\phi}. 
\end{equation}

\subsection{Asymptotic Anaysis for $\phi_c \ll 1$}
The function $F(\phi;\phi_c)$ is given by, 
\begin{equation}
    F(\phi;\phi_c) = \frac{g(\phi_c)}{\sqrt{2(\cos\phi - \cos\phi_c)}} - \int_{\phi}^{\phi_c} \frac{g'(u) \, \mathrm{d}u}{\sqrt{2(\cos\phi - \cos u)}}. 
\end{equation}
In the small $\phi_c$ limit this becomes, 
\begin{equation}
     F(\phi;\phi_c) \sim \frac{g(\phi_c)}{\sqrt{\phi_c^2 - \phi^2}} - \mathcal{C}(\phi;\phi_c)
\end{equation}
where $\mathcal{C}(\phi;\phi_c)$ is a correction integral given by, 
\begin{align}
    \mathcal{C}(\phi;\phi_c) = \int_{\phi}^{\phi_c} \frac{g'(u)\,\mathrm{d}u}{\sqrt{u^2 - \phi^2}}. 
\end{align}

The quantity of primary interest is the dimensionless evaporative flux given by,
$$
\frac{\phi_c^2 F(\phi;\phi_c)}{I(\phi_c)}
$$
If we let $S(\tphi) = 1/\sqrt{1-\tphi^2}$ then
\begin{align*}
    \phi_c^2 F(\phi;\phi_c) &\sim  n_1 \phi_c  + n_2 {\phi_c^2}\log\phi_c+ n_3 {\phi_c^2},\\
    n_1 &= S(\tphi),\\
    n_2 &= -(1/\pi)S(\tphi),\\
    n_3 &= ((1-\log 2)/\pi)S(\tphi) - \mathcal{C}(\tphi)
\end{align*}
while    
\begin{equation}
    I(\phi_c) = \int_0^{\phi_c} F\sin\phi \, \mathrm{d}\phi \sim \int_0^{\phi_c} F\phi \,\mathrm{d}\phi = \int_0^1 \phi_c^2 F \cdot \tphi \, \mathrm{d}\tphi .
\end{equation}
This can then be expanded as 
\begin{equation}
    I(\phi_c) \sim d_1 \phi_c + d_2 \phi_c^2 \log \phi_c + d_3 \phi_c^2,
\end{equation}
where
\begin{align}
    d_1 &= \int_0^1 \frac{x}{\sqrt{1-x^2}}\,\mathrm{d}x = 1\\
    d_2 &= \frac{-1}{\pi}\int_0^1 \frac{x}{\sqrt{1-x^2}}\,\mathrm{d}x =  \frac{-1}{\pi}\\
    d_3 &=\frac{1-\log 2}{\pi} - \Pi(\tphi=1),
\end{align}
and $\Pi(x) = \int_0^x s {\mathcal{C}}(s) \, ds$. 
The RHS of Equation \eqref{eqn:COFapp} then looks like  
\begin{align}
\frac{\phi_c^2 F(\phi;\phi_c)}{4I(\phi_c;\phi_c)} \sim J_0 + \phi_c J_1    
\end{align}
where
\begin{align}
    J_0 &= \frac{1}{4\sqrt{1-\tphi^2}},\\
    J_1 &= \frac{\Pi(1)}{4\sqrt{1-\tphi^2}} - \frac{\mathcal{C}(\phi;\phi_c)}{4}
\end{align}
Thus, corrections to the local evaporative flux enter at $O(\phi_c)$. A similar analysis for $\tilde{\eta}$ shows that corrections enter at $O(\phi_c^2)$. For Equation \eqref{eqn:COFapp} to balance, there must be an $O(\phi_c)$ correction to the fluid velocity which we conclude is due to the evaporative flux being modified as outlined in the main text. 
\end{document}